\documentclass[12pt]{article}
\usepackage{amsmath}

\usepackage{scicite}
\usepackage{times}
\usepackage{graphicx}

\newenvironment{sciabstract}{\begin{quote} \bf}
{\end{quote}}

\newcounter{lastnote}
\newenvironment{scilastnote}{\setcounter{lastnote}{\value{enumiv}}\addtocounter{lastnote}{+1}\begin{list}{\arabic{lastnote}.}
{\setlength{\leftmargin}{.22in}}
{\setlength{\labelsep}{.5em}}}
{\end{list}}

\def\cpl{ Chem.\ Phys.\ Lett.\ }

\def\jcp{ J.\ Chem.\ Phys.\ }

\def\nat{ Nature (London) }

\def\pra{ Phys.\ Rev.\ A }

\def\prl{ Phys.\ Rev.\ Lett.\ }

\usepackage[small]{caption2}

\begin{document}

\title{Optical pumping and vibrational cooling of molecules}
\author{Matthieu Viteau,$^{1}$ Amodsen Chotia,$^{1}$ Maria Allegrini,$^{1,2}$
Nadia Bouloufa,$^{1}$ \and Olivier Dulieu,$^{1}$ Daniel Comparat,$^{1}$
Pierre Pillet$^{1\ast }$ \\
\\
$^{1}$\ {\normalsize {Laboratoire Aim\'{e} Cotton, CNRS, Univ Paris-Sud, B%
\^{a}t. 505, 91405 Orsay, France}}\\
{\normalsize {$^{2}$ CNISM, Dipartimento di Fisica, Universit\`{a} di Pisa, Largo 
Pontecorvo 3, 56127 Pisa, Italy}}\\
{\normalsize {$^{\ast }$To whom correspondence should be addressed; E-mail:
pierre.pillet@lac.u-psud.fr} }}
\maketitle

\baselineskip14pt

\begin{sciabstract}
The methods producing cold molecules from cold atoms tend to leave molecular ensembles with substantial residual internal energy. For instance, Cs$_2$ molecules initially formed via photoassociation of cold Cs atoms are in several vibrational levels, $v$, of the electronic ground state. Here we apply a broadband femtosecond laser that redistributes the
vibrational population in the ground state via a few electronic excitation - spontaneous emission cycles. The laser
pulses are shaped to remove the excitation frequency band of the $v=0$ level, preventing re-excitation from that state. We observe a fast and efficient accumulation, $\sim70\%$ of the initially detected molecules, in the lowest vibrational level, $v=0$, of the singlet electronic state. The validity of this incoherent depopulation pumping method is very general and opens exciting prospects for laser cooling and manipulation of molecules.
\end{sciabstract}

In the last twenty years atomic
physics has made enormous strides, with laser cooling and the achievement
of atomic Bose-Einstein condensation. Similar advances are expected with cold molecules, involving applications for instance in molecular clocks, tests on fundamental physical constants or quantum computing. Thus, the preparation of dense molecular samples in the ground state at low temperatures offers exciting prospects in both physics and chemistry \cite{2004EPJD...31..149D,Krems,DulieuJPB2006}. 

An important step in the field of cold  molecules has been the
demonstration of a method for producing translationally cold samples of ground-state Cs$_{2}$ molecules via
photoassociation of cold Cs atoms \cite{1998PhRvL..80.4402F}. This result has been quickly followed by the
elaboration of various methods to prepare cold molecular samples.
Methods that start with pre-formed molecules, usually in the lowest vibrational level, access translational temperature down to a few millikelvins\cite{1998Natur.395..148W,1999PhRvL..83.1558B,2003Sci...302.1940E,GuptaM.jp993560x,2003PhRvA..67d3406R}.
Accessing even lower than these temperatures presents a major challenge. Cold molecules
in the micro- or nano-kelvin temperature range,
can only be achieved starting with cold atoms using
collisional processes such as photoassociation in a thermal atomic cloud \cite{1998PhRvL..80.4402F}, Feshbach magneto-association in atomic Bose-Einstein condensates \cite{PhysRevLett.93.123001}, or
three-body collisions in an atomic Fermi sea to prepare molecular Bose-Einstein condensates \cite {2003Sci...302.2101J}. 
These methods of producing (translationally) cold molecules from cold atoms, however, lead to the production of vibrationally excited molecules, ie. with significant residual internal energy. For additional applications of cold molecules \cite{2004EPJD...31..149D,Krems,DulieuJPB2006}, the challenge is therefore to prepare and control molecules in the ground vibrational and rotational state.

Various experimental schemes can favor the formation of cold
molecules in their lowest vibrational level. In a quantum gas, the adiabatic transfer of population (STIRAP: STImulated Raman Adiabatic Passage) from a high ro-vibrational level towards a lower one has been achieved recently for molecules formed by magneto-association \cite%
{2007PhRvL..98d3201W}. In a cold thermal gas, a fraction of cold ground state Rb-Cs molecules, initially formed by photoassociation, has been prepared into the lowest vibrational level, $v=0$, with a rate about 500$\,$s$^{-1}$ by transferring $\sim6\%$ of the population of a given high vibrational level into $v = 0$ \cite{2005PhRvL..94t3001S}. A few $v=0$ cold ground-state
potassium dimers have also been observed by using a two-photon process
for photoassociation \cite{1999PhRvL..82..703N}, but several
other vibrational levels are populated as well. For further applications, what is needed is a molecular analogue of optical pumping of atoms, to realize vibrational laser
cooling, which would transfer all the populations of the
different vibrational levels into the lowest one.

Several theoretical approaches have been proposed to favor spontaneous emission towards the
lowest ro-vibrational level: for instance, use of an external cavity 
\cite{2007PhRvL..99g3001M} or controlled interplay of coherent laser fields and spontaneous emission 
through quantum interferences between different
transitions  \cite{1999FaraDisc,1993JChPh..99..196B,2001PhRvA..63a3407S}. As in these latter coherent control propositions,
our approach uses a shaped pulsed laser, but it is based on an incoherent process of depopulation pumping with a train of several identical weak femtosecond laser pulses. More closely related to our work is the proposition of using a tailored incoherent broadband light source for rotational cooling of molecular ions \cite%
{2002PhRvL..89q3003V,2004JPhB...37.4571V}.

Here, we report transfer of populations
from an ensemble of vibrational levels of cold Cs$_{2}$
molecules prepared in the electronic ground-state via photoassociation, into $v=0$. The main idea is to use a broadband laser tuned to
the transitions between the different vibrational levels, which we label $v_{\text{X}}$ and $v_{\text{B}}$, of the singlet-ground-state, X, and of an electronically excited state, B. The absorption - spontaneous emission cycles lead, through optical pumping, to a redistribution of the vibrational population into the ground
state (eq. \ref{equ:Equ1})
\begin{equation}
	\text{Cs}_{2}(v_{\text{X}})+h\nu \longrightarrow \text{Cs}_{2}(v_{\text{B}})\overset{%
	decay}{\longrightarrow }\text{Cs}_{2}(v^{'}_{\text{X}})
\label{equ:Equ1}
\end{equation}
 where ideally $v^{'}_{\text{X}}<v_{\text{X}}$ in order to realize vibrational cooling. The broadband character of the laser permits repetition of the pumping process from multiple vibrational states. By removing the laser frequencies corresponding to the excitation of the $v_{\text{X}}=0$ level, we make it impossible to pump molecules out of this level, thus making $v_{\text{X}}=0$ a dark state. As time progresses the absorption - spontaneous emission cycle described by Eq.(\ref{equ:Equ1}), leads to an accumulation of the molecules in the $v_{\text{X}}=0$ level. We thereby realize vibrational laser cooling.

\begin{figure}[ht!]
	\centering
	\resizebox{1\textwidth}{!}{	
		\includegraphics*[41mm,218mm][183mm,267mm]{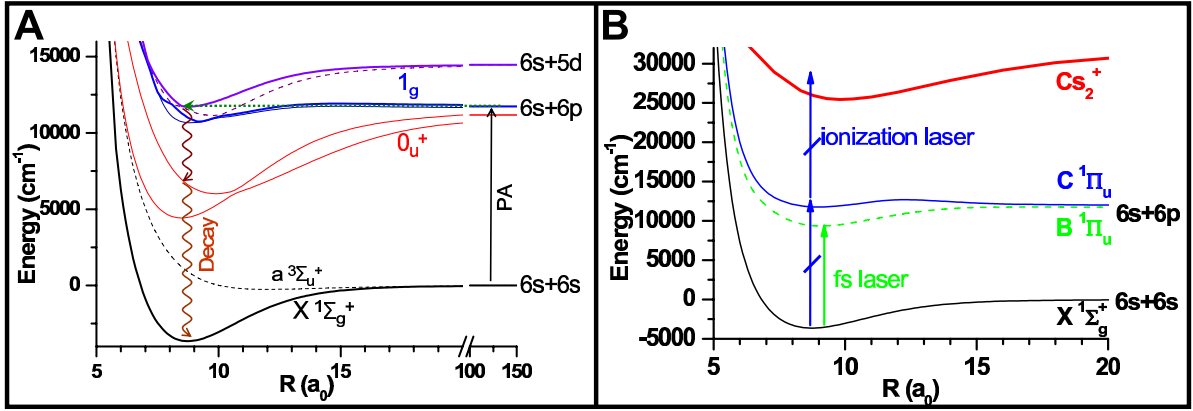}}
	\caption{ Relevant schematic molecular potential curves of the cesium dimer, converging towards the
dissociation limits 6s+6s, 6s+6p, and 6s+5d (for clarity, the fine structure is not labelled). \textbf{A}, Photoassociation of cold
atoms and formation of cold molecules. The cw laser (PA) is tuned $\sim 1$ cm$^{-1}$ below
the atomic transition $6s_{1/2}\longrightarrow 6p_{3/2}$. For the potentials of 1$_{g}$ symmetry,
long-range radial wavefunction is coupled to short range radial wavefunction
by internal coupling of the potentials \cite{2001PhRvL..86.2253D}. The ground-state molecules, X$%
^{1}\Sigma _{g}^{+}$, are formed by a spontaneous emission cascade via the 
$0_{u}^{+}$ potentials. \textbf{B}, REMPI\ ionization process via the C$^{1}\Pi _{u}$
state by the pulsed dye laser, and electronic transition X$^{1}\Sigma _{g}^{+}$ towards B$^{1}\Pi _{u}$ induced by the femtosecond laser.}
	\label{fig:fig1}
\end{figure}

\begin{figure}[ht!]
	\centering
	\resizebox{0.7\textwidth}{!}{	
		\includegraphics*[26mm,166mm][166mm,265mm]{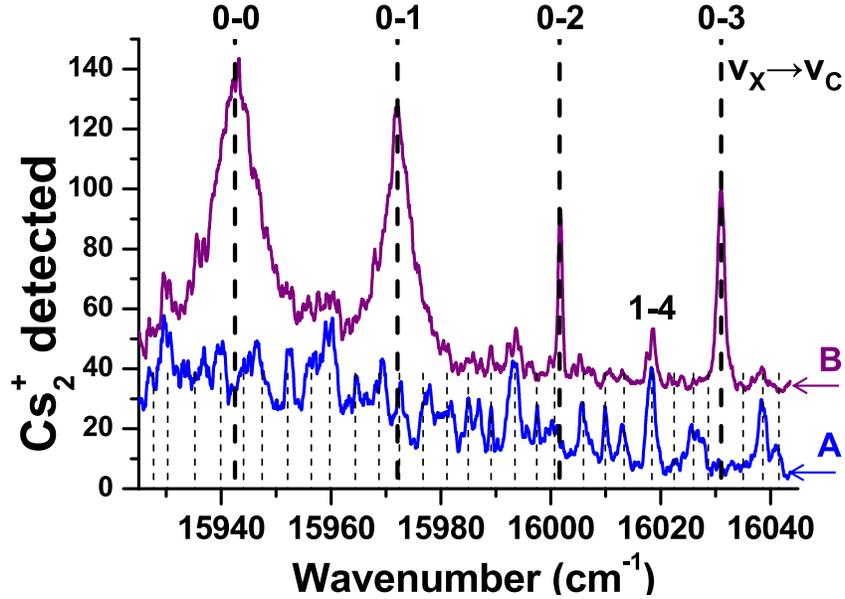}}
	\caption{Cs$_{2}^{+}$ ion spectra. \textbf{A}, Spectrum without the shaped laser pulse. The
spectrum has a background due to other REMPI
processes which do not 
mask the resonance lines. The dashed lines indicate the positions of all the resonances 
for vibrational transitions 
between the ground state, X$^{1}\Sigma _{g}^{+}$ ($v_{\text{X}}=0$ to $7$), and the electronically
excited C$^{1}\Pi _{u}$ state ($v_{\text{C}}$). \textbf{B}, Spectrum with the
shaped laser pulse applied continuously, offset by 40 ions for higher visibility. The observed transitions from $v_{\text{X}}=0$  correspond to $v_{\text{C}}=$ 0, 1, 2 and 3. Their broadening correspond to the saturation of the resonance in the REMPI process. Most of the lines present in spectrum \textbf{A} are greatly reduced while the $v_{\text{X}}=0$ lines grow more intense. The resonance labeled 1-4 indicates imperfect depopulation of v$_X$=1 due to the roughness of the shaping.}
	\label{fig:fig2}
\end{figure}

In our experiment, the formation of cold molecules is achieved in a Cs vapor-loaded Magneto-Optical Trap (MOT), via photoassociation \cite{1998PhRvL..80.4402F}. Two colliding cold atoms
resonantly absorb a photon with a frequency tuned slightly ($\sim$1 cm$^{-1}$) below the atomic 6s$_{1/2}$-6p$_{3/2}$ transition to create a molecule in an excited electronic state. The photoassociated molecules decay by spontaneous emission into stable vibrational levels of the molecular ground state, $\text{X}^{1}\Sigma _{g}^{+}$ (see Fig.\ref{fig:fig1}A). They are then detected by  Resonantly Enhanced Multiphoton Ionization (REMPI). In contrast with previous studies \cite{1998PhRvL..80.4402F}, the REMPI frequency is tuned to ionize deeply bound vibrational levels of the X state, through the excited $\text{C}^{1}\Pi _{u}$ molecular state (see Fig.\ref{fig:fig1}B). The complete mechanism for the formation of cold molecules in the singlet ground state is under study but the most probable scenario is given in Fig.\ref{fig:fig1}A. Photoassociation is achieved using a cw Titanium:Sapphire laser (intensity 300 W.cm$^{-2}$) pumped by an Argon-ion laser. The REMPI detection uses a pulsed dye laser (wavenumber $\sim$16000 cm$^{-1}$, spectral bandwidth 0.3 cm$^{-1}$) pumped by the second harmonic of a pulsed Nd:YAG laser (repetition rate 10Hz,
duration 7ns). The formed Cs$_{2}^{+}$ ions are detected using a pair of microchannel plates through a time-of-flight mass spectrometer. In the experimental spectrum, obtained by scanning the REMPI laser wavelength (Fig.\ref{fig:fig2}A), we assigned the observed lines to known transitions from the ground state levels $v_{\text{X}}=1$ to 7 to various levels of the C state \cite{1982JChPh..76.4370R}. In this first step, no molecules in the vibrational level, $v_{\text{X}}$ = 0, are detected. The present low REMPI resolution does not provide the capability of analyzing the rotational population of the molecules.

\begin{figure}[ht!]
	\centering
	\resizebox{1\textwidth}{!}{	
		\includegraphics*[8mm,202mm][181mm,256mm]{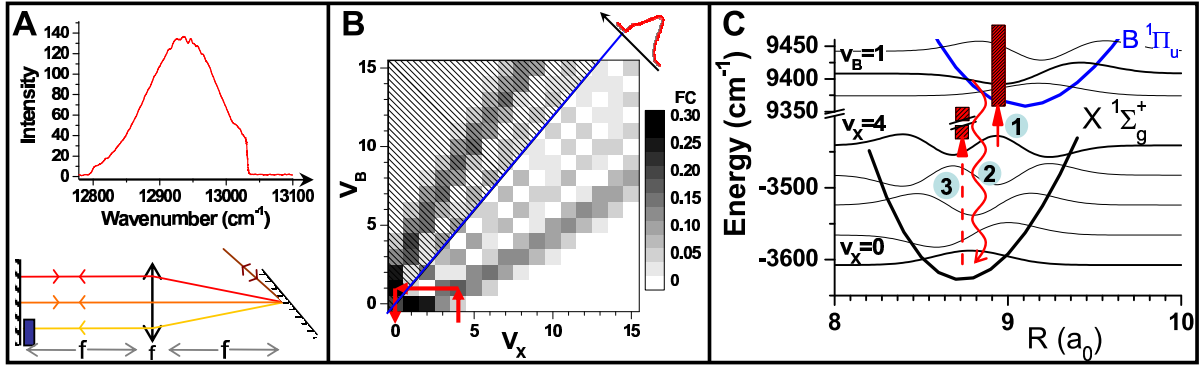}}		
	\caption{Optical pumping scheme with a shaped femtosecond laser pulse. \textbf{A}, Upper part, the line gives the laser spectrum profile with shaping (shown also in \textbf{B}), close to a gaussian profile sharply cut off at 13030 cm$^{-1}$. Lower part, 4-f shaping arrangement (from right to left: grating; cylindrical lens $f=500$mm; blocker; mirror). \textbf{B}, Condon parabola indicating the importance (level of grey) of the Franck-Condon (FC) factors (square of the wavefunction overlap), corresponding to the relative transition probabilities from $v_{\text{X}}$ to $v_{\text{B}}$. The diagonal line corresponds to the shaped laser cutoff frequency (13030 cm$^{-1}$). The hatched area cannot be accessed in the presence of the blocker. \textbf{C}, Optical pumping scheme and vibrational wavefunctions. The vertical black boxes indicate the spectral bandwidth of the laser. In \textbf{B} and \textbf{C}, arrows indicates the optical pumping for $v_{\text{X}}=4$ molecules. The most probable
optical pumping scheme is to reach $v_{\text{X}}=0$ through excitation into $v_{\text{B}}=1$. Step (\textbf{1}), excitation towards $v_{\text{B}}=1$. Step (\textbf{2}), spontaneous decay to $v_{\text{X}}=0$. Step (\textbf{3}), molecules in $v_{\text{X}}=0$ are trapped. The incoherent dark state formed by the laser pulse shaping does not allow the excitation from $v_{\text{X}}=0$ to any $v_{\text{B}}$ level.} 
	\label{fig:fig3}
\end{figure}

To achieve vibrational cooling, we applied a broadband femtosecond mode-locked laser (repetition rate 80\
MHz, pulse duration 100 fs, standard deviation-gaussian bandwidth 54\ cm$^{-1}$, average intensity of $50$ mW/cm$^{2}$, central wavelength 773 nm or wavenumber 12940 cm$^{-1}$) tuned to the electronic transitions from X$^{1}\Sigma _{g}^{+}(v_{%
\text{X}})$ to $\ $B$^{1}\Pi _{u}(v_{\text{B}})$ (Fig.\ref{fig:fig1}B).  
Without shaping the femtosecond laser pulses, we observe a modification of the vibrational distribution, which we interpret as a transfer of population between vibrational levels as indicated by Eq.(\ref{equ:Equ1}). The relative strength of the transitions between the vibrational levels of the  X-B states are given by the Franck-Condon factors (Fig.\ref{fig:fig3}B). If we consider for instance a molecule in $v_{\text{X}}=4$, the most probable excitation is to $v_{\text{B}}=1$, which decays as in Eq.(\ref{equ:Equ1}) with a partitioning ratio of about 30\% to $v^{'}_{\text{X}}=0$, and 70\% distributed essentially among $v^{'}_{\text{X}}=3$, 4 and 5. To control the optical pumping of the molecules, we shaped the femtosecond laser pulses by suppressing the frequencies above 13030\ cm$^{-1}$ that could induce electronic excitation from $v_{\text{X}}=0$ (see Fig.\ref{fig:fig3}A and hatched area in Fig.\ref{fig:fig3}B). We used a home-built shaper with a diffraction grating (1800 lines per mm) after which high frequencies of the laser beam are screen out (see lower part of Fig.\ref{fig:fig3}A). After a few cycles of absorption of
laser light and spontaneous emission, considering the populations in the observed vibrational levels ( $v_{\text{X}}=0-10$), a large fraction, $65\% \pm10\%$, of the molecules are accumulated in the lowest vibrational level, $v_{\text{X}}=0$.

\begin{figure}[ht!]
	\centering
	\resizebox{1\textwidth}{!}{	
	\rotatebox{-90}{
		\includegraphics*[1mm,2mm][52mm,136mm]{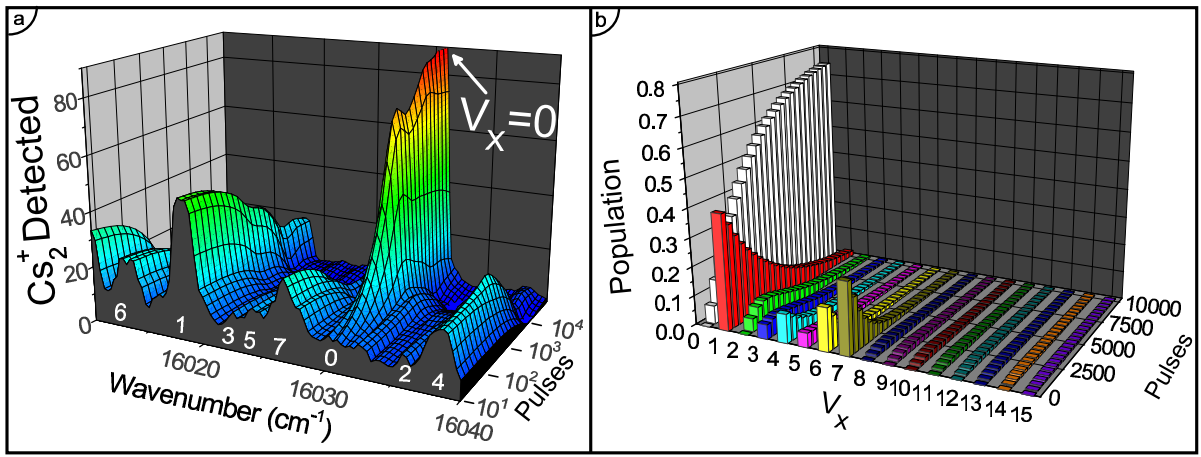}}}
	\caption{Temporal evolution (pulse separation 12.5ns) of the populations in the different vibrational levels of the ground state. Due to our weak laser intensity, the excitation probability is only $0.1\%$ for a single pulse. \textbf{A}, Experimental population versus the number of applied femtosecond pulses, smoothed from 5 spectra (similar to Fig.\ref{fig:fig2} but taken after controlled number of pulses). The frequencies correspond to transitions from $v_{\text{X}}=0-7$ to $v_{\text{C}}$ levels (white label). \textbf{B}, Theoretical simulation where we represent the temporal evolution of $v_{\text{X}}=0$ to $15$ starting with initial conditions close to the experimental ones.}
	\label{fig:fig4}
\end{figure}

On application of the shaped laser pulses, the resonance lines corresponding to transition from $v_{\text{X}}=0$ to $v_{\text{C}}=0$ to 3, emerged strongly in the REMPI spectra  (Fig.\ref{fig:fig2}B). The intensity of
the lines indicates efficient transfer of the molecules into the lowest
vibrational level $v_{\text{X}}=0$.
By controlling the number of femtosecond laser pulses with an acousto-optic modulator, we analyzed the time dependence of the optical pumping scheme (Fig.\ref{fig:fig4}A). 
At the weak laser intensities applied here, the transfer of population into the $v_{%
\text{X}}=0$ level is almost completed after an exposure of the sample to 5000 pulses over 60$\mu$s. Taking into account the
efficiency of the detection, the detected ion signal
corresponds to about one thousand molecules in the $v_{\text{X}}=0$ level in the MOT area,
and thus to a formation rate of $v_{\text{X}}=0$ molecules of more than 10$^{5}$ per
second, which represent roughly 1\% of the atomic loading flux in the MOT.

We have modeled the optical pumping process using the experimentally known X$^{1}\Sigma _{g}^{+}$ and B$^{1}\Pi _{u}$ potential curves 
\cite{1985JChPh..82.5354W,1989CPL...164..419D}. In our perturbative regime,
the excitation probabilities are proportional to the laser spectral
density at the transition frequencies.  The lifetime of the electronically excited
state B, $\sim 15$ ns, is close to the 12.5 ns repetition period of the femtosecond laser, leaving negligible accumulation of coherence in the sample from pulse to pulse \cite{2003OptCo.215...69F}. We then assumed in our rate equation model, an instantaneous spontaneous decay. 
The model shows
that the vibrational population ($v_{\text{X}}$) proceeds by random walk, mostly through low vibrational levels,
until reaching the $v_{\text{X}}=0$ level. More than 70 \% of the total population is transferred into the $v_{\text{X}}=0$ level (Fig.\ref{fig:fig4}B), when we start from a distribution of vibrational levels close to
the experimental one. The simulation shows that the limitation of the efficiency of the mechanism is in the optical pumping towards higher vibrational levels. Nevertheless, the simulation demonstrates that for instance increasing the bandwidth of the laser would reduce this detrimental pumping and would increase the population in $v_{\text{X}}=0$. The theoretical model agrees well with the
data in Fig.\ref{fig:fig4}A. Furthermore, it indicates that only about five absorption - spontaneous emission cycles, corresponding to $\sim$5000 laser pulses, are necessary for a molecule to be transferred into the $v_{\text{X}}=0$ level. This small number of cycles does not significantly modify the temperature of the molecular sample. The theoretical simulation takes into account the rotational levels and demonstrates, for the experiment, the possibility to achieve rotational cooling for an adapted shaping, accurate enough to resolve the rotational structure.

The method, optical pumping of diatomic molecules, using a shaped broadband source, is expected to be generally applicable to most molecular sample experiments which present a distribution of population of the low vibrational levels in the ground state. The efficiency will depend on the transition strengths between the different vibrational levels of the considered electronic states, but it could be optimized with a suitable shaping. The optical pumping should not be limited to cold samples of molecules prepared via photoassociation of cold atoms, but should be applicable also to other case such as molecules in a molecular beam. Broadband shaped optical pumping could also be used as a repumping laser in laser manipulation of atoms and molecules, opening prospects in laser cooling of new species \cite{1996JChPh.104.9689B}.
\bigskip

\bibliographystyle{Science}

\begin{scilastnote}
\item
The authors thank Thomas F. Gallagher for helpful discussions during the
redaction of this article.\ They acknowledge fruitful debates with F.\
Masnou-Seeuws, E.\ Luc-Koenig, A.\ Crubellier and B.\ Chatel about the applications at
the frontier of the ultracold and ultrafast fields.\ M.A. thanks the EC-Network EMALI. This work is supported
by the "Institut Francilien de Recherche sur les Atomes Froids" (IFRAF).\
The laser cooling development is performed in the frame of the ``Agence Nationale de la Recherche" (ANR) Grant CORYMOL.
\end{scilastnote}

\clearpage

\end{document}